\documentclass[a4paper,fleqn,usenatbib]{mnras}

\usepackage[T1]{fontenc}
\usepackage{ae,aecompl}

\usepackage{graphicx}	
\usepackage{amsmath}	
\usepackage{amssymb}	

\title[The AMBRE Project: chemical evolution models]{The AMBRE Project: chemical evolution models for the Milky Way thick and thin discs}

\author[Grisoni et al.]{V. Grisoni$^1$\thanks{E-mail: grisoni@oats.inaf.it}, E. Spitoni$^1$, F. Matteucci$^{1, 2, 3}$, A. Recio-Blanco$^4$, P. de Laverny$^4$,\newauthor M. Hayden$^4$, $\check{\text{S}}$. Mikolaitis$^{4,5}$, and C.C. Worley$^6$\\
 $^1$ Dipartimento di Fisica, Sezione di Astronomia, Universit\`a di Trieste, via G.B. Tiepolo 11, I-34131, Trieste, Italy \\  
 $^2$ I.N.A.F. Osservatorio
  Astronomico di Trieste, via G.B. Tiepolo 11, I-34131, Trieste,
  Italy\\
 $^3$ I.N.F.N. Sezione di Trieste, via Valerio 2, 34134 Trieste, Italy\\
 $^4$ Universit\`e C$\hat{o}$te d'Azur, Observatoire de la C$\hat{o}$te d'Azur, CNRS, Laboratoire Lagrange, Bd de l'Observatoire, CS 34229,\\06304 Nice Cedex 4, France\\
 $^5$ Institute of Theoretical Physics and Astronomy, Vilnius University, Saul$\dot{e}$tekio al. 3, 10257, Vilnius, Lithuania\\
 $^6$ Institute of Astronomy, University of Cambridge, Madingley Road, Cambridge, CB3 0HA, UK
}

\begin{document}
\date{Accepted . ; in original form xxxx}

\pagerange{\pageref{firstpage}--\pageref{lastpage}} \pubyear{xxxx}

\maketitle

\label{firstpage}

\begin{abstract}
We study the chemical evolution of the thick and thin discs of the Galaxy by comparing detailed chemical evolution models with recent data from the AMBRE Project.
The data suggest that the stars in the thick and thin discs form two distinct sequences with the thick disc stars showing higher [$\alpha$/Fe] ratios. We adopt two different approaches to model the evolution of thick and thin discs. In particular, we adopt: i) a two-infall approach where the thick disc forms fast and before the thin disc and by means of a fast gas accretion episode, whereas the thin disc forms by means of a second accretion episode on a longer timescale; ii) a parallel approach, where the two discs form in parallel but at different rates. By comparing our model results with the observed [Mg/Fe] vs. [Fe/H] and the metallicity distribution functions in the two Galactic components, we conclude that the parallel approach can account for a group of $\alpha$-enhanced metal rich stars present in the data, whereas the two-infall approach cannot explain these stars unless they are the result of stellar migration. In both approaches, the thick disc has formed on a timescale of accretion of 0.1 Gyr, whereas the thin disc formed on a timescale of 7 Gyr in the solar region. In the two-infall approach a gap in star formation between the thick and thin disc formation of several hundreds of Myr should be present, at variance with the parallel approach where no gap is present.
\end{abstract}


\section{Introduction}

In recent years, many spectroscopic surveys and projects have been developed in order to study the formation and evolution of the Milky Way, such as for example Gaia-ESO (Gilmore et al. 2012), APOGEE (Majewski et al. 2015) and the AMBRE Project (de Laverny et al. 2013). Furthermore, the arrival of Gaia data is enhancing the value of these surveys. For instance, the Gaia/RVS data will provide abundances data for several tenths of millions of stars (Recio-Blanco et al. 2016).
\\In this way, detailed stellar abundances of stars in the Milky Way can be measured. In particular, the latest observational data reveal a clear distinction between the abundance patterns of the thick and thin disc stars, especially for the $\alpha$-elements. In fact, Gaia-ESO data (Recio-Blanco et al. 2014; Rojas-Arriagada et al. 2017), APOGEE data (Hayden et al. 2015) and AMBRE data (Mikolaitis et al. 2017) indicate two distinct sequences corresponding to thick and thin disc stars, and the presence of these two sequences still has to be interpreted in terms of Galactic chemical evolution models.
\\As pointed out in Matteucci (2012), Galactic chemical evolution models have passed through different phases that can be summarized as follows: i) Serial approach (e.g. Matteucci and Francois 1989); ii) Parallel approach (e.g. Ferrini et al. 1992, Pardi et al. 1995, Chiappini 2009); iii) Two-infall approch (e.g. Chiappini et al. 1997, Romano et al. 2010): iv) Stochastic approach (e.g. Argast et al. 2000, Cescutti 2008).
\\In the serial approach, one assumes that the halo, thick and thin-disc form in sequence. In this framework, the thick disc is simply a later phase relative to the halo and the thin disc is a later phase relative to the thick disc. In the parallel approach, the various Galactic stellar components start forming at the same time but evolve in parallel at different rates. The two-infall model belongs to the serial approach, but it assumes that the halo-thick disc formed out of a completely independent gas accretion episode relative to the thin disc. This latter formed out of different extragalactic gas on a much longer timescale. In the stochastic approach, the early phases of the evolution are characterized by inhomogeneities of the interstellar medium (ISM), in the sense that the early supernovae (SNe) pollute only nearby regions and the mixing is not efficient.
\\In this paper, we will model the thick and thin disc evolution by adopting both the two-infall and the parallel approach.
\\The parallel approach was first introduced by Ferrini et al. (1992) and Pardi et al. (1995). In their model, they consider the three phases 
(halo, thick and thin discs) to evolve separately and in parallel. However, a limitation of this model is that the three phases are connected to one another through the infalling gas and this fact prevents to obtain a good agreement with the stellar metallicity distribution functions in the three phases. In fact, the three observed metallicity distribution functions are different, indicating that each component cannot have formed out of gas shed by the other two (see Matteucci, 2001). An advantage of their approach is that they can explain the observed spread in the data and the observed overlapping in metallicity of stars belonging to different components. A more recent parallel approach was suggested by Chiappini (2009), Anders et al. (2017). The main difference between Chiappini (2009) and Pardi et al. (1995) is that in Chiappini's approach the thick and thin disc evolutions are completely disentangled.
\\On the other hand, the two-infall approach assumes two main infall episodes: during the first one, the halo-thick disc formed, whereas the second one gave rise to the thin disc. On this line, Chang et al. (1999) applied the two-infall model of Chiappini et al. (1997) to the thick and thin discs, although the data at that time were much less and sparse. In the original two-infall model the thick disc was assumed to form fast on a timescale no longer than 2 Gyr and it was considered together with the halo.  Micali et al. (2013) extended the two-infall model into a three infall model where the formation of the thick disc was assumed to have occurred by means of a gas  accretion episode totally independent from the episodes forming the halo and the thin disc. In their model, the thick disc formed faster than the thin disc and on a timescale of 1 Gyr. They were able to reproduce the stellar metallicity distribution functions of the thick and thin discs. However, for what concerns the [$\alpha$/Fe] ratios the available data were too sparse to identify different trends between the thick and thin disc stars. Recently, on the basis of the data of Adibekyan et al. (2012) Haywood et al. (2015) studied the evolution of the thick disc and concluded that the star formation history was uniform throughout the thick disc. They also concluded that the thick disc did not form inside-out in the first 3-5 Gyr of the evolution of the Galaxy. Later on, Haywood et al. (2016) by considering APOGEE data concluded that there was a quenching in the star formation at the end the thick phase. Kubryk et al. (2015) suggested instead that the thick disc is the result of stellar migration: they concluded that the thick disc is the early part of the Milky Way disc. They explained the sequences of [$\alpha$/Fe] ratios of the thick and thin discs by analyzing the data of Bensby et al. (2014). Masseron and Gilmore (2015)  by studying the APOGEE data concluded that the majority of thick disc stars formed earlier than the thin disc ones and that the star formation rate in the thick disc was more efficient than in the thin disc.
\\The aim of this paper is to reproduce the chemical characteristics of the thick and thin disc stars as observed by the most recent data of the AMBRE Project (Mikolaitis et al. 2017). The AMBRE abundances come from high resolution data, similarly to those of Adibekyan et al. (2012), but they belong to a much larger sample. The AMBRE resolution is also higher than APOGEE data. In order to study these data, we test the two-infall and parallel scenarios, by means of improved and updated Galactic chemical evolution models. Our chemical models are based on the two-infall model (Chiappini et al. 1997, Romano et al. 2010) revisited and applied to the thick and thin discs and a new parallel model adopting two one-infall models for the thick and thin discs, respectively. In this way, the evolution of the thick and thin discs are completely disentangled. 
\\The paper is organized as follows. In Section 2, we present the data which have been used to make a comparison with the predictions of our chemical evolution models.
In Section 3, we describe the chemical evolution models adopted. In Section 4, we show the comparison between model predictions and observations. Finally, Section 5 summarizes our results and conclusions.

\section{Observational data}

The observational data used in this work for comparison with the chemical evolution models are issued from the AMBRE Project (de Laverny et al. 2013). We remind that AMBRE has been defined in order to homogeneously determine stellar atmospheric parameters and chemical abundances for the archived spectra of the ESO spectrographs for Galactic archaeology purposes. Up to now, more than 200,000 HR spectra (including several repeats for several stars) have already been analysed. The corresponding atmospheric parameters have been derived thanks to the MATISSE algorithm (Recio-Blanco et al. 2006) and a large grid of FGKM synthetic spectra (de Laverny et al. 2012). In the present paper, we have adopted the magnesium and iron chemical abundances presented in Mikolaitis et al. (2017). These abundances have been derived owing to an automatic line-fitting technique for the AMBRE FEROS and HARPS spectra, which have been previously parametrized by Worley et al. (2012) and de Pascale et al. (2014), respectively. We also point out the AMBRE sample is not complete and is characterized by different observational biases inherent to the content of the ESO archive. The present sample consists in 4,666 individual slow-rotating stars, most of them ($\sim$ 11\%) being dwarfs of the solar neighborhood for which accurate Mg and Fe are available. All of these targets have been classified owing to their Mg and Fe properties into five different Galactic components (thin and thick discs, metal-poor high/low $\alpha$, and metal-rich high $\alpha$, MRHA hereafter). We remind that it has been possible to conduct such a chemical labelling thanks to the small uncertainties of the derived abundances and also because magnesium is one of the best specy to separate the two Galactic discs (Mikolaitis et al., 2014).
\\Finally, we have recently analysed the kinematical and dynamical properties of the stars in the AMBRE catalogue (Hayden et al. in prep.). Those stars are also part of the TGAS catalogue (Brown et al. 2016, Lindegren et al. 2016) included in the first data release of the Gaia mission (Prusti et al. 2016). Thanks to the Gaia precise astrometry, we have been able to derive reliable orbital parameters for the stars using the galpy code (Bovy et al. 2015). In particular, the perigalacticon points of the orbits, derived by Hayden et al. (private communication) will be used in this paper.

\section{The models}

The chemical evolution models adopted here are:
\begin{itemize}
\item the two-infall model (Chiappini et al. 1997, Romano et al. 2010) revisited and applied to the thick and thin discs, and
\item a new parallel model adopting two one-infall models for the thick and thin discs, respectively.
\end{itemize}

\subsection{The two-infall model}

The two-infall model adopted here is a revision of the model developed by Chiappini et al. (1997), Romano et al. (2010). This model assumes that the Galaxy forms as a result of two main infall episodes. During the first one, the thick disc formed, whereas during the second one a much slower infall of gas, delayed with respect to the first one, gives rise to the thin disc. Here, we do not take into account the evolution of the halo, but we focus on the evolution of the thick and thin discs. The origin of the gas in the infall episodes is extragalactic and its composition is assumed to be primordial. The Galactic thin disc is approximated by several independent rings, 2 kpc wide, without exchange of matter between them whereas the evolution of the thick disc is fixed with radius.
\\The basic equations that describe the time evolution of $G_i$, namely the mass fraction of the element $i$ in the gas, are (see Matteucci, 2012):
\begin{align} \label{eq_01}
\begin{split}
& \dot G_i(r,t)= -\psi(r,t) X_i(r,t)\\
& + \int\limits^{M_{Bm}}_{M_L(t)}\psi(r,t-\tau_{m})Q_{mi}(t-\tau_{m})\phi(m)dm\\
& + A\int\limits^{M_{BM}}_{M_{Bm}}\phi(m)\cdot\left[\int\limits_{\mu_{m}}^{0.5}f(\mu)\psi(r,t-\tau_{m2})Q_{mi}(t-\tau_{m2})d\mu\right]dm\\
& + (1-A)\int\limits^{M_{BM}}_{M_{Bm}}\psi(r,t-\tau_{m})Q_{mi}(t-\tau_{m})\phi(m)dm\\
& + \int\limits^{M_{U}}_{M_{BM}}\psi(r,t-\tau_{m})Q_{mi}(t-\tau_{m})\phi(m)dm\\
& + \dot G_i(r,t)_{inf}
\end{split}
\end{align}
As described in Matteucci (2012), the first term in the right hand side of Eq.\,\eqref{eq_01} represents the rate at which the chemical elements are subtracted from the ISM to be included into stars, whereas the various integrals represent the rate of restitution of matter from the stars into the ISM. In particular,
\begin{itemize}
\item The first integral represents the material restored by stars in the mass range $M_L(t)-M_{Bm}$, where $M_L(t)$ is the minimum mass dying at the time t and its minimum value is $\simeq 0.8 M_{\bigodot}$ with a lifetime corresponding to the age of the Universe. The stellar lifetime of a star of mass $m$ is defined as $\tau_m$, and $\tau_{m_2}$ is the lifetime  of the secondarty star of those binary systems giving rise to SNe Ia (see next). In fact, the clock to the explosion of these systems is given by the lifetime of the secondary star in the single degenerate model (Matteucci and Greggio 1986). The term $Q_{mi}$ is the new and already present fraction of an element $i$ which is restored into the ISM by a star of mass $m$.
\item The second integral corresponds to the contribution of Type Ia SNe, as first introduced by Matteucci and Greggio (1986). Here, the rate is calculated by assuming the single-degenerate model for the progenitor of these SNe, namely a C-O WD plus a red giant companion. The extremes of this integral correspond to the minimum mass $M_{Bm}$ and the maximum mass $M_{BM}$ allowed for the whole binary systems giving rise to Type Ia SNe. The minimum mass $M_{Bm}$ is set equal to $3 M_{\bigodot}$, in order to ensure that both the primary and the secondary star would be massive enough to allow the WD to reach the Chandrasekhar mass $M_{Ch}$, after accretion from the companion. The maximum mass is fixed by the requirement that the mass of each component cannot exceed $M_{up}= 8 M_{\bigodot}$, which is the assumed maximum mass giving rise to a C-O WD, and so $M_{BM}= 16 M_{\bigodot}$. The function $f(\mu_B)$ describes the distribution of the mass ratio of the secondary of the binary system ($\mu_B = \frac{M_2}{M_B}$). The parameter A represents the fraction of binary systems with the right properties to give rise to SNe Ia and it is assumed A=$0.035$. The time $\tau_{m2}$ is the lifetime of the secondary star in the binary system giving rise to a SN Ia, and represents the clock of the system in the single degenerate scenario.
In Matteucci et al. (2009), it has been demonstrated that this scenario is equivalent to the double degenerate one as the effects on Galactic chemicevolution are concerned.
\item The third integral refers to single stars with masses in the range $M_{Bm}-M_{BM}$ (namely, $3 M_{\bigodot} - 16 M_{\bigodot}$). They can be either stars ending their lives as C-O WDs or as Type II SNe (those with M $>$ $M_{up}$ which is assumed to be 8 $M_{\bigodot}$).
\item The fourth integral represents the material restored by core-collapse SNe.
\end{itemize}
The last term in the equation is the gas accretion rate. In particular, the gas infall law is described as:
\begin{align} \label{eq_02}
\dot G_i(r,t)_{inf}=A(r)(X_i)_{inf}e^{-\frac{t}{\tau_1}}+B(r)(X_i)_{inf}e^{-\frac{t-t_{max}}{\tau_2}},
\end{align}
where $G_i(r,t)_{inf}$ is the infalling material in the form of element $i$ and $(X_i)_{inf}$ is the composition of the infalling gas which is assumed to be primordial. The parameter t$_{max}$ is the time for the maximum mass accretion onto the disc and roughly corresponds to the end of the thick disc phase. The parameters $\tau_1$ and $\tau_2$ are the timescales for mass accretion in the thick and thin disc components, respectively: they are the e-folding times of the mass accretion law and represent the times at which each component accumulated roughly half of its mass. These timescales are free parameters of the model and they are constrained mainly by comparison with the observed metallicity distribution of long-lived stars in the solar vicinity. The quantities $A(r)$ and $B(r)$ are two parameters fixed by reproducing the present time total surface mass density in the solar neighborhood as taken from Nesti and Salucci (2013). In particular, this is equal to 65 M$_{\odot}$pc$^{-2}$ for the thin disc and 6.5 M$_{\odot}$pc$^{-2}$ for the thick disc. Other studies suggest slightly different values for the total local surface mass density (Bovy and Rix 2013; Zhang et al. 2013; McKee et al. 2015): we tested also these values and found negligible difference in the results. What really matters here is the ratio between the total surface mass density of the thick and thin discs, which is considered to be 1:10.
\\The star formation rate (SFR) is the Schmidt-Kennicutt law (Kennicutt 1998a):
\begin{equation} \label{eq_03}
\psi(r,t)=\nu \sigma_{gas}^k(r,t),
\end{equation}
where $\sigma_{gas}$ is the surface gas density, $k$ is the law index and $\nu$ the star formation efficiency (i.e. the star formation rate per unit mass of gas). The star formation efficiency is assumed to become zero when the surface gas density goes below a critical threshold $\sigma_{th}$ (Kennicutt 1998a,b; Martin and Kennicutt 2001).
\\The initial mass function (IMF) can be parameterized as a power law of the following kind:
\begin{equation} \label{eq_04}
\phi(m)=am^{-(1+x)},
\end{equation}
generally defined in the mass range of 0.1-100 M$_{\bigodot}$, where $a$ is the normalization constant derived by imposing that:
\begin{equation} \label{eq_05}
\int_{0.1}^{100} m\phi(m) dm=1.
\end{equation}
In particular, we adopt the Kroupa et al. (1993) IMF that corresponds to:
\begin{align} \label{eq_06}
\begin{split}
& x=0.3 \text{ for } M \le 0.5 M_{\bigodot}\\
& x=1.2 \text{ for } 0.5M_{\bigodot} < M \le 1.0 M_{\bigodot}\\
& x=1.7 \text{ for } M > 1.0 M_{\bigodot}
\end{split}
\end{align}

\subsection{The parallel model}

Secondly, we consider the possibility of abandoning a sequential scenario like the one of the two-infall, in favour of a picture which treats the thick disc and the thin disc stars as two truly distinct evolutionary phases, which start at the same time but evolve independently. In the light of these considerations, we develop two distinct one-infall models: one for the thick disc and the other for the thin disc.
\\As in the previous model, the material accreted by the Galactic discs comes mainly from extragalactic sources, and the basic equation is the same seen before, i.e. Eq. \,\eqref{eq_01}.
\\Since this model assumes two distinct infall episodes, the gas infall is described as:
\begin{equation} \label{eq_07}
(\dot G_i(r,t)_{inf})|_{thick}=A(r)(X_i)_{inf}e^{-\frac{t}{\tau_1}},
\end{equation}
\begin{equation} \label{eq_08}
(\dot G_i(r,t)_{inf})|_{thin}=B(r)(X_i)_{inf}e^{-\frac{t}{\tau_2}},
\end{equation}
for the thick disc and for the thin disc, respectively. The quantities $A(r)$ and $B(r)$ and the parameters $\tau_1$ and $\tau_2$ have the same meaning as discussed for Eq. \,\eqref{eq_02}. Actually, the exponential form is similar to the case of the two-infall model, but the novelty introduced here concerns the fact that the infall rate of the thick and thin discs are now totally disentangled. In fact, as mentioned above, we want to treat the thick and the thin disc as two truly distinct evolutionary phases. 
\\For the SFR and the IMF, the functional forms are the same of the two-infall model (Eq. \,\eqref{eq_03} and Eq. \,\eqref{eq_06}, respectively).

\subsection{Nucleosynthesis prescriptions}

The nucleosynthesis prescriptions and the implementation of the yields in the model are fundamental ingredients for chemical evolution models. In this work, we adopt the same nucleosynthesis prescriptions of model 15 of Romano et al. (2010), where an exhaustive description of the adopted yields can be found.
\\For the computation of the stellar yields, one has to distinguish between different mass ranges as well as single stars versus binary systems:
\begin{itemize}
\item low and intermediate mass stars (0.8 $M_{\bigodot}$-8 $M_{\bigodot}$), which are divided into single stars and binary systems which can give rise to Type Ia SNe,
\item massive stars (M $>$ 8 $M_{\bigodot}$).
\end{itemize}
Single stars in the mass range 0.8 $M_{\bigodot}$-8 $M_{\bigodot}$ contribute to the Galactic chemical enrichment through planetary nebula ejection and quiescent mass loss along the giant branch. They enrich the interstellar medium mainly in He, C, and N, but they can also produce some amounts of $^7$Li, Na and $s$-process elements. For these stars, which end their lives as white dwarfs, the adopted nucleosynthesis prescriptions are from Karakas (2010).
\\Type Ia SNe are considered to originate from carbon deflagration in C-O white dwarfs in binary systems. These stars contribute a substantial amount of iron (0.6 $M_{\bigodot}$ per event) and non negligible quantities of Si and S. They also contribute to other elements, such as O, C, Ne, Ca, Mg and Mn, but in negligible amounts with respect to the masses of such elements ejected by Type II SNe. The adopted nucleosynthesis prescriptions are from Iwamoto et al. (1999).
\\Massive stars with masses M $>$ 8 $M_{\bigodot}$ are the progenitors of Type II, Ib and Ic SNe: if the explosion energies are much higher than 10$^{51}$ erg, hypernova events can occur (SNe Ic). For these stars, the adopted nucleosynthesis prescriptions are from Kobayashi et al. (2006) for the following elements: Na, Mg, Al, Si, S, Ca, Sc, Ti, Cr, Mn, Co, Ni, Fe, Cu and Zn. As for the He and CNO elements, we consider the results of Geneva models for rotating massive stars (see Romano et al. 2010). However, for Mg which is one the relevant element in this study we adopted yields multiplied by a factor 1.2 in order to obtain a better agreement with the data. It is well known, in fact, that Mg yields have been  underestimated in many nucleosynthesis studies (see Fran\c cois et al. 2004 for a discussion of this point), and although the most recent ones have improved nonetheless the Mg production in massive stars is still underestimated.

\begin{table*}
\caption{Input parameters for the best chemical evolution models. 2IM corresponds to the two-infall model, whereas 1IMT and 1IMt correspond to the one-infall models for the thick and thin discs, respectively. In the second column, we show the adopted initial mass function. In the third and fourth columns, there are the star formation efficiencies for the thick and thin discs, respectively. In the fifth and sixth columns, we give the timescales for mass accretion in the thick and thin discs, respectively. Finally, in last column, we show the adopted threshold in the star formation rate.}
\label{tab_01}
\begin{center}
\begin{tabular}{c|cccccccccc}
  \hline
\\
 Model & IMF &$\nu_1$& $\nu_2$ &$\tau_1$& $\tau_2$ & $\sigma_{th}$\\
&  &[Gyr$^{-1}$]&[Gyr$^{-1}$]& [Gyr]&[Gyr]&[M$_{\odot}$pc$^{-2}$]\\
\\
\hline

2IM & Kroupa & 2& 1  & 0.1 &7 & 7 \\

 \hline

1IMT & Kroupa & 2& - & 0.1 & - & - \\

 \hline

1IMt & Kroupa & - & 1  & - &7 & 7 \\

 \hline

\end{tabular}
\end{center}
\end{table*}

\section{Results}

Good models of Galactic chemical evolution should reproduce the majority of the observational features and always a number of observational constraints that is larger than the number of free parameters. In this work, the observational constraints considered are the following ones:
\begin{itemize}
\item Abundance patterns of the most common chemical elements, in particular the [Mg/Fe] vs. [Fe/H] abundance pattern recently observed by the AMBRE Project;
\item Metallicity distribution function of long-lived stars belonging to the thick and thin disc components as recently observed by the AMBRE Project;
\item Solar abundances;
\item Present-time SFR;
\item Present-time Type Ia and Type II supernova rates.
\end{itemize}

\begin{figure*}
\includegraphics[scale=0.5]{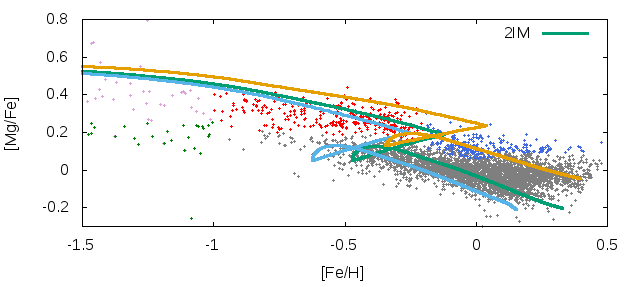}
\includegraphics[scale=0.5]{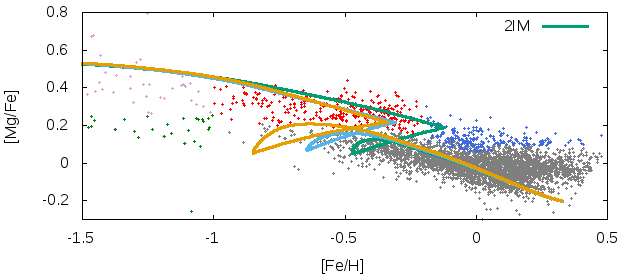}
\includegraphics[scale=0.5]{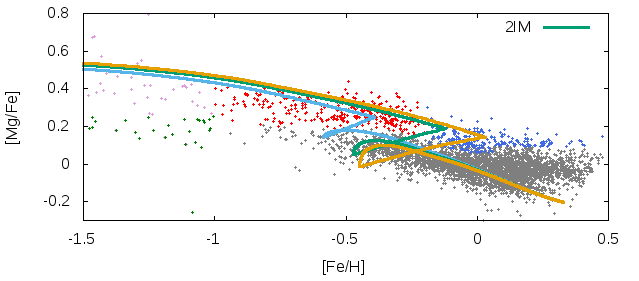}
\includegraphics[scale=0.5]{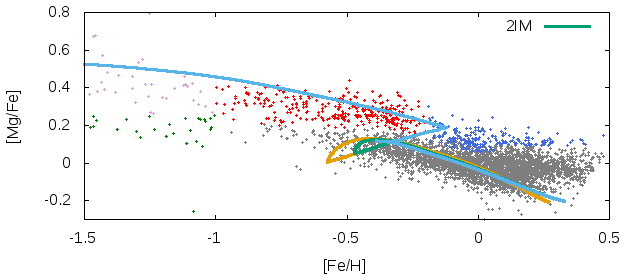}
 \caption{Predicted and observed [Mg/Fe] vs. [Fe/H] in the solar neighborhood in the case of the two-infall model. The data are from the AMBRE Project and the different Galactic components are plotted: thin disc (grey dots), thick disc (red dots), metal-rich high-$\alpha$ sequence (blue dots), metal-poor low-$\alpha$ sequence (green dots), metal poor high-$\alpha$ sequence (magenta). \textit{Upper left panel:} the effect of varying the IMF. The 2IM with Kroupa et al. (1993) IMF (green line) is compared with a model with Scalo (1986) IMF (light-blue line) and Salpeter (1955) IMF (orange line). \textit{Lower left panel:} the effect of varying the star formation efficiency of the thick disc. The 2IM with $\nu_1=2$ Gyr$^{-1}$ (green line) is compared with a model with $\nu_1=1$ Gyr$^{-1}$ (light-blue line) and $\nu_1=3$ Gyr$^{-1}$ (orange line). \textit{Upper right panel:} the effect of varying the timescale of the thick disc. The 2IM with $\tau_1=0.1$ Gyr (green line) is compared with a model with $\tau_1=0.5$ Gyr (light-blue line) and $\tau_1=1$ Gyr (orange line). \textit{Lower right panel:} the effect of varying the threshold. The 2IM with $\sigma_{th}=7$ M$_{\bigodot}$ $pc^{-2}$ (green line) is compared with a model with $\sigma_{th}=4$ M$_{\bigodot}$ $pc^{-2}$ (light-blue line) and $\sigma_{th}=10$ M$_{\bigodot}$ $pc^{-2}$ (orange line).}
 \label{fig_00}
\end{figure*}

Our best models have been selected after running several numerical simulations by varying one at the time the most important input parameters. The input parameters  of the best models are summarized in Table 1. In the first column, we give the names of the models: 2IM corresponds to the two-infall model, whereas 1IMT and 1IMt correspond to the one-infall models for the thick and thin discs, respectively. In the second column, we show the adopted IMF. In the third and fourth columns, there are the star formation efficiencies for the thick and thin discs, respectively. In the fifth and sixth columns, we give the timescales for mass accretion in the thick and thin discs, respectively. Finally, in last column, we show the adopted threshold in the SFR.
\\In Fig. \ref{fig_00}, we show the effect of varying one at the time the most important input parameters of the two-infall model.
\\In the upper left panel of Fig. \ref{fig_00}, we show the effect of varying the IMF. We can see the prediction of the two-infall model in the solar vicinity for which we consider Kroupa et al. (1993) IMF, compared to the case with Scalo (1986) IMF and Salpeter (1955) IMF. We can see that Scalo (1986) IMF predicts too few massive stars, and so the corresponding track in the [Mg/Fe] vs. [Fe/H] lies below the data. On the other hand, Salpeter (1955) IMF predicts too many massive stars, and so the corresponding track in the [Mg/Fe] vs. [Fe/H] is above the data.
\\In the lower left panel of Fig. \ref{fig_00}, we show the effect of varying the star formation efficiency of the thick disc. We can see the prediction of the two-infall model in the solar vicinity for which the star formation efficiency of the thick disc is $\nu_1=$ 1, 2 and 3 Gyr$^{-1}$. We can see that a higher star formation efficiency implies a higher track in the [Mg/Fe] vs. [Fe/H], even if the effect is less strong than in the case of varying the IMF. Furthermore, we can see that a higher star formation efficiency means a more rapid evolution for the thick disc, and this extends the range of [Mg/Fe] values for the thick disc stars.
\\In the upper right panel of Fig. \ref{fig_00}, we show the effect of varying the timescale for mass accretion in the thick disc. We can see the prediction of the two-infall model in the solar vicinity for which the timescale of the thick disc is $\tau_1=$ 0.1, 0.5 and 1 Gyr. We can see that a shorter timescale for the thick disc formation extends the range of [Mg/Fe] values for the thick disc stars, because the evolution is more rapid.
\\In the lower right panel of Fig. \ref{fig_00}, we show the effect of varying the threshold in the SFR. We can see the prediction of the two-infall model in the solar vicinity for which the threshold is $\sigma_{th}=$ 4, 7 and 10 M$_{\bigodot}$ $pc^{-2}$. The gap in the model prediction is due to the assumed threshold in the star formation process. Varying the threshold means varying the extension of this gap, and the gap increases with increasing threshold.
\\On the other hand, the two parameters regarding the star formation efficiency and the timescale for mass accretion in the thin disc are well constrained by reproducing the G-dwarf distribution and correspond to $\nu_2=1$ Gyr$^{-1}$ and $\tau_2=7$ Gyr, as found in previous studies (Chiappini et al. 1997; Romano et al. 2010).
\\Similarly, we chose also the various input parameters for the one-infall models, as summarized in Table 1.
\\In the following, we focus on our best models and we show the predictions concerning the abundance patterns, the metallicity distribution functions of the thick and thin discs, the solar abundances, the star formation history and the supernova rates.

\begin{figure*}
\centering
\includegraphics[scale=0.6]{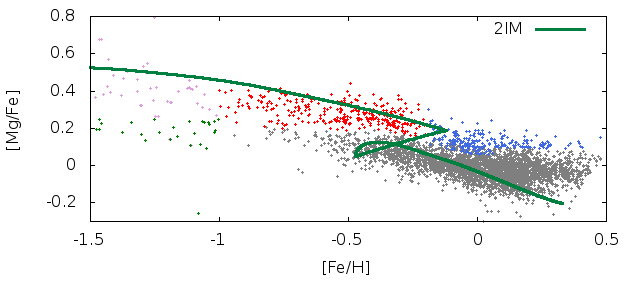}
 \caption{Predicted and observed [Mg/Fe] vs. [Fe/H] in the solar neighborhood in the case of the two-infall model. The data are from the AMBRE Project and the different Galactic components are plotted: thin disc (grey dots), thick disc (red dots), metal-rich high-$\alpha$ sequence (blue dots), metal-poor low-$\alpha$ sequence (green dots), metal poor high-$\alpha$ sequence (magenta). The predictions are from model 2IM (green line).}
 \label{fig_01}

\centering
\includegraphics[scale=0.8]{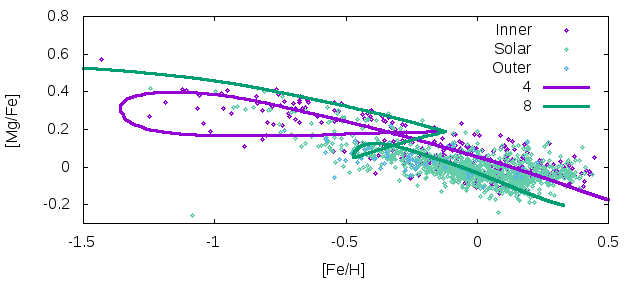}
 \caption{Predicted and observed [Mg/Fe] vs. [Fe/H]. The data are color-coded according to their guiding radius (inner, solar, outer) and the predictions are from model 2IM (both at 8 kpc and at 4 kpc).}
 \label{fig_01b}
\end{figure*}

\subsection{Abundance patterns}

\begin{figure*}
\centering
\includegraphics[scale=0.6]{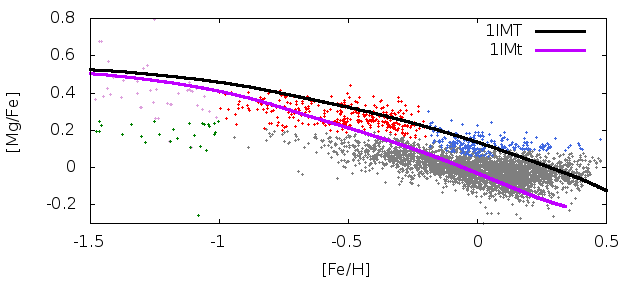}
 \caption{Same as Fig. 1, but the predictions are for models 1IMT for the thick disc (black line) and 1IMt for the thin disc (magenta line).}
 \label{fig_02}
\end{figure*}

\begin{figure*}
\includegraphics[scale=0.35]{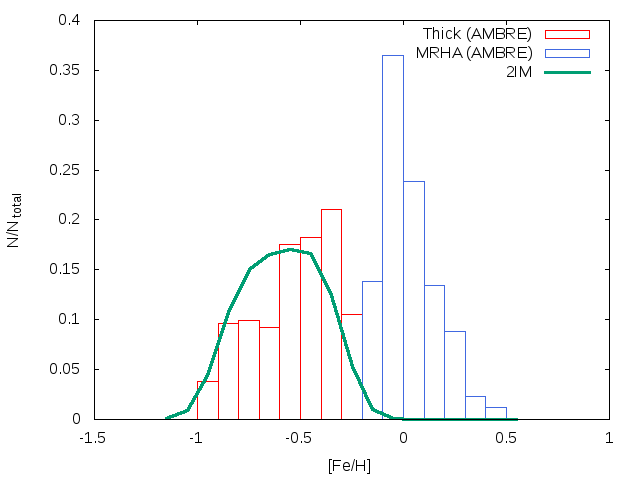}
\includegraphics[scale=0.35]{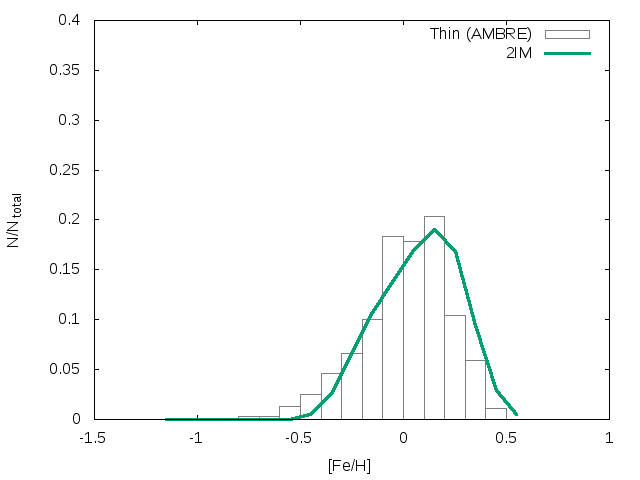}
\caption{{\it Left panel:} Predicted and observed metallicity distribution function of the thick disc in the case of the two-infall model. The data are from the AMBRE Project: MDF of thick disc stars (red) and MDF of MRHA stars (blue). The predictions are from model 2IM (green line). We notice that the two-infall model cannot reproduce the MRHA stars, as explained in the text. {\it Right panel:} Predicted and observed metallicity distribution function of the thin disc in the case of the two-infall model. The data are from the AMBRE Project: MDF of thin disc stars (grey). The predictions are from model 2IM (green line).}
 \label{fig_03}

\includegraphics[scale=0.35]{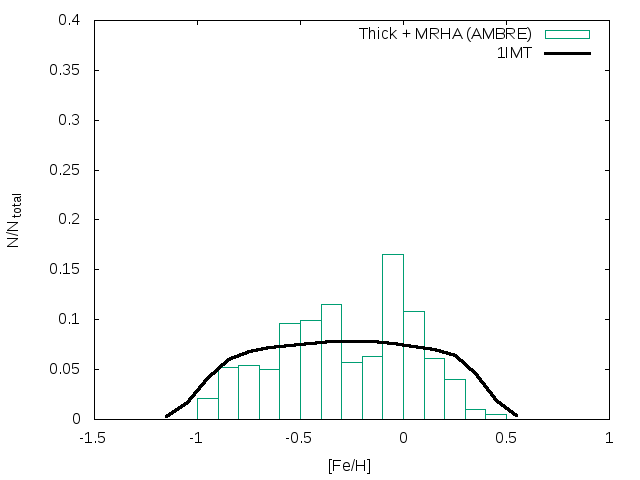}
\includegraphics[scale=0.35]{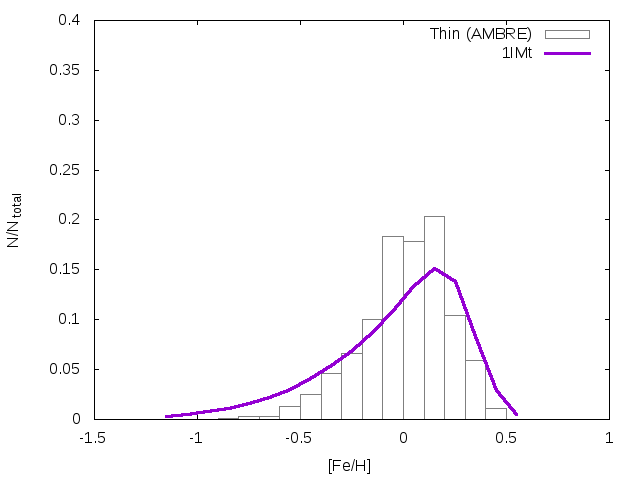}
\caption{Same as Fig. 4, but the predictions are for models 1IMT for the thick disc (black line) and 1IMt for the thin disc (magenta line). We notice that in this case in the left panel the MDF of thick plus MRHA stars is considered, since the one-infall model of the thick disc can reproduce also the MRHA stars.}
 \label{fig_04}
\end{figure*}

The first observational constraint considered is the [Mg/Fe] vs. [Fe/H] relation, as observed by  the AMBRE Project.
\\In Fig. \ref{fig_01}, we show the predicted and observed [Mg/Fe] vs. [Fe/H] in the solar neighborhood in the case of the two-infall model 2IM. The model predicts an overabundance of Mg relative to Fe almost constant until [Fe/H]$<$-1.5 dex and then for [Fe/H]$\sim$-1.5 dex the trend shows a slight decrease due to the delayed explosion of Type Ia SNe. This behaviour of the abundance patterns of $\alpha$-elements such as Mg is well-interpreted in terms of the time-delay model  (see Matteucci 2001; 2012): the time-delay refers to the delay of iron ejection from Type Ia SNe relative to the faster production of $\alpha$-elements by core-collapse SNe. The effect of the delayed iron production is to create an overabundance of $\alpha$-elements relative to iron at low [Fe/H] values, and a continuous decline of the [$\alpha$/Fe] ratio until the solar value is reached. A peculiar feature of the 2IM is that at [Fe/H]$\sim$-0.2 dex we have a gap of 700 Myr duration due to the assumed threshold in the star formation process, which marks the transition between the thick and thin disc phases, where the star formation stops until the gas density in the thin disc reaches the threshold. A similar gap was found in the two-infall model of Chiappini et al. (1997), but between the halo-thick and thin disc phases, whereas here we focus on the disc (thick plus thin). It is worth noting that the 2IM model does not reproduce the metal rich $\alpha$-enhanced stars (MRHA) (blue dots). If they are thick disc stars, this model cannot reproduce them since the thick disc does not extend far enough in [Fe/H]. However, these stars have shorter Rp (perigalacticon) values and we could suppose that we see them now in the solar neighborhood thanks to stellar migration from the inner thin disc.
\\In Fig. \ref{fig_01b}, we show the predictions of the 2IM also in a slice that is more internal than the solar neighborhood, i.e. 4 kpc from the Galactic center. What differ from the 2IM for the solar neighborhood are the timescale of the thin disc (which is shorter and at 4kpc is equal to 0.5 Gyr in the framework of an inside-out scenario) and the surface mass density (which changes with radius and at 4 kpc is equal to 303 M$_{\odot}$pc$^{-2}$ as in Nesti and Salucci 2013). According to the 2IM at 4 kpc, a large loop is evident in the abundance pattern: this is due to the fact that the initial gas infall in the inner disc is more efficient than in the solar ring, and this produces a large [Fe/H] dilution between the thick and thin disc formation. In this case, the track in the abundance pattern is higher at higher metallicities and the MRHA stars can be fitted. Hence, these stars could be interpreted as stars that have migrated from the inner thin disc.
\\Then, in Fig. \ref{fig_02}, we show the predicted and observed [Mg/Fe] vs. [Fe/H] in the solar neighborhood in the case of the parallel models 1IMT and 1IMt for the thick and thin discs, respectively. In the case of a parallel model, the thick and thin discs are treated as two truly distinct evolutionary phases and so we have two distinct tracks in the abundance pattern. Both the tracks show an overabundance of Mg relative to Fe almost constant until [Fe/H]$<$-1.5 dex and then the trends show the decrease due to the delayed explosion of Type Ia SNe.
However, the [$\alpha$/Fe] ratios in the thick disc stars are higher than in the thin disc, as a consequence of the assumed faster evolution of the thick disc (a timescale of the order of 0.1 Gyr, as shown in Table 1). Indeed, the fit to the data requires that the formation of the thick disc occurs on shorter timescales than the formation of the thin disc. In fact, a fast SFR and a short timescale of gas accretion are required to form the thick disc, whereas a much slower SFR and longer accretion timescale (7 Gyr, see Table 1) are necessary to reproduce the features of the thin disc.
In Figure 2 it is evident that with the parallel model we are able to reproduce the MRHA stars as metal rich thick disc stars, since in the parallel approach the thick disc can extend up to high [Fe/H], at variance with the 2IM model.

\subsection{Metallicity distribution functions}

A fundamental constraint that we have to analyze is the metallicity distribution function (MDF), both in the thick and thin discs. Thanks to our models, we are able to create two different MDFs, one for the thick and one for the thin disc.
\\First, let us consider the case of the two-infall model. Since the two-infall model predicts a gap in the star formation between the thick and thin discs, we consider as stars of the thick disc all those formed before the gap in the star formation and as stars of the thin disc all those formed afterwards.
\begin{itemize}
\item In the left panel of Fig. \ref{fig_03}, we show the predicted and observed MDF of the thick disc in the case of the two-infall model: the data are from the AMBRE Project and the predictions are from model 2IM. From the data, we can see that the metallicity of the thick disc goes from [Fe/H]$\sim$-1.0 dex to [Fe/H]$\sim$-0.2 dex, and the mean value is $<$[Fe/H]$>$ $\sim$-0.5 dex. The 2IM well reproduces the observations. In fact, we have a tail of metal-poor stars and the same relative number for higher-metallicity stars, and the relative number of stars with [Fe/H]$\sim$-0.5 dex predicted by the model is equal to $\sim$ 15$\%$. Therefore, as regards to the thick disc, the model is good from the chemical point of view, as it predicts a relative number of stars in agreement with the observations, in the right [Fe/H] range. However, we notice that the 2IM model cannot reproduce the MRHA stars, because in the two-infall approach the thick disc phase does not extend so far in metallicity.
\item In the right panel of Fig. \ref{fig_03}, we show the predicted and observed MDF of the thin disc in the case of the two-infall model: the data are from the AMBRE Project and the predictions are from model 2IM. From the data, we can see that the metallicity of the thin disc goes from [Fe/H]$\sim$-0.5 dex to [Fe/H]$\sim$0.5 dex, and the mean value is around the solar metallicity. Also in the case of the thin disc, the model is in good agreement with observations.
\end{itemize}
Secondly, let us consider the case of the parallel model.
\begin{itemize}
\item In the left panel of Fig. \ref{fig_04}, we show the predicted and observed MDF of the thick disc in the case of the parallel model: the data are from the AMBRE Project and the predictions are from model 1IMT. To have a good agreement with the data, we have to consider the MDF of thick plus MRHA stars, which can be reproduced in the framework of the 1IMT model. In fact, the MDF predicted by model 1IMT is very broad and includes also the MRHA stars.
\item In the right panel of Fig. \ref{fig_04}, we show the predicted and observed MDF of the thin disc in the case of a parallel model: the data are from the AMBRE Project and the predictions are from model 1IMt. Concerning the thin disc, the model is in quite good agreement with the data of the thin disc stars.
\end{itemize}

\subsection{Solar abundances}

In Table 2, the solar abundances predicted by the models 2IM and 1IMt are compared with observations of Grevesse et al. (2007).
\\The abundances are expressed as 12+log(X/H). These abundances correspond to the composition of the interstellar medium at the time of the formation of the Sun, 4.5 Gyr ago. Since we assume a Galactic lifetime of 13.7 Gyr, we have calculated the solar abundances at 9.2 Gyr after the Big Bang. Given the uncertainties, we can say that the predictions are in reasonable agreement with the observations. However,  our predicted solar O is always  a bit overestimated and the good agreement with the Mg solar abundance is due to the fact that we did increase the Mg yields from massive stars by multiplying them by a factor 1.2.

\begin{table}
\caption{Solar abundances (in dex).}
\label{tab_02}
\begin{tabular}{c|ccc}
  \hline
\\
 Elem. & Observations & 2IM & 1IMt \\
\\

 \hline

O & 8.66$\pm$0.05 & 8.95 & 8.96 \\

\hline

Mg & 7.53$\pm$0.09 & 7.58 & 7.59 \\

 \hline

Si & 7.51$\pm$0.04 & 7.71 & 7.72 \\

 \hline

S & 7.14$\pm$0.04 & 7.34 & 7.36 \\

 \hline

Fe & 7.45$\pm$0.05 & 7.64 & 7.66 \\

 \hline

\end{tabular}
\end{table}

\subsection{Star formation history}

\begin{figure}
\includegraphics[scale=0.4]{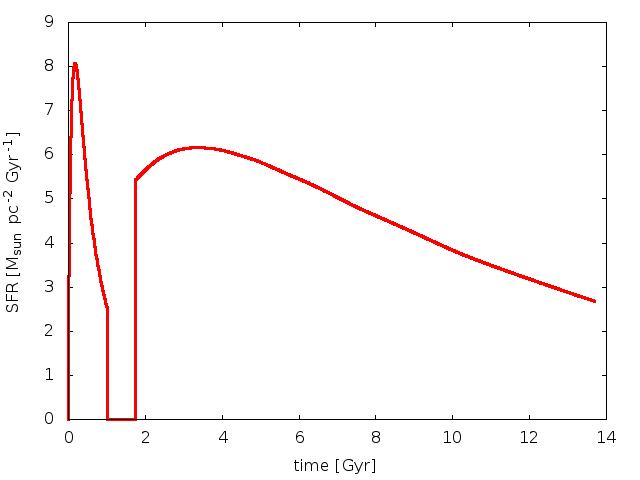}
 \caption{Temporal evolution of the SFR, as predicted by model 2IM.}
 \label{fig_07}

\includegraphics[scale=0.4]{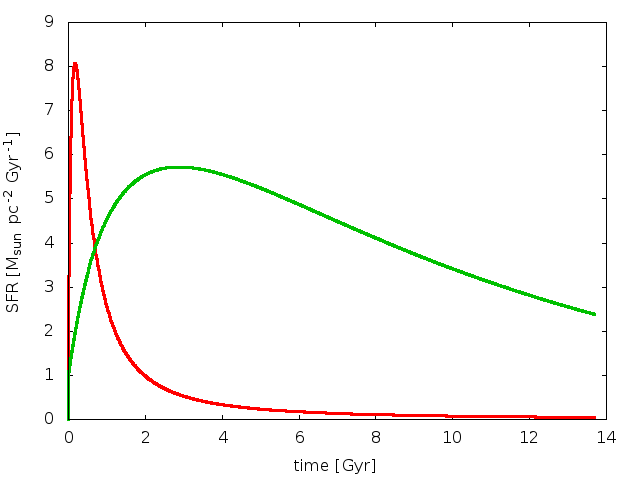}
 \caption{Temporal evolution of the SFR. \textit{Red line:} prediction of model 1IMT. \textit{Green line:} prediction of model 1IMt.}
 \label{fig_08}
\end{figure}

In Fig. \ref{fig_07}, we show the SFR versus time as predicted by model 2IM. As we can see, the SFR is higher during the thick disc phase, while it is lower during the thin disc formation. A peculiar feature of this plot is the gap in the SFR between the thick disc and the thin disc formation. This is clearly due to the fact that the star formation in the thin disc occurs only after a density of 7 M$_{\bigodot}$ pc$^{-2}$ has been accumulated. 
This gap was already predicted by Chiappini et al. (1997) for the 2IM model applied to the halo-thick disc and thin disc. The duration of the predicted gap between the thick and thin disc formation here is roughly 1 Gyr.
An important constraint is represented by the present time SFR in the solar vicinity, and according to Bovy (2017) the present time SFR as measured with Gaia is:
$$\psi_0 \sim 1.7 \text{ M}_{\bigodot} \text{ pc}^{-2} \text{ Gyr}^{-1},$$
with an e-folding time of $\sim$ 7 Gyr. The value predicted by our best model is:
$$\psi_0=2.7 \text{ M}_{\bigodot} \text{ pc}^{-2} \text{ Gyr}^{-1},$$
in good agreement with observations.
\\In Fig. \ref{fig_08}, we show the SFR versus time as predicted by models 1IMT and 1IMt for the thick and thin discs, respectively. The two SFRs are now separate, since we have assumed a distinct evolution for the two components of the disc, and there is no gap in star formation between the thick and thin disc phases. The star formation history of the thick disc is very different from the star formation history of the thin disc. In fact, the star formation history of the thick disc is peaked at earlier times, because it forms more rapidly than the thin disc (in fact, the thick disc has a higher star formation efficiency and a shorter timescale of formation with respect to the thin disc). On the other hand, the SFR of the thin disc has a peak shifted to a later time and has still an active star formation at the present time. The value predicted for the present time SFR in the solar vicinity by model 1IMt is:
$$\psi_0=2.4 \text{ M}_{\bigodot} \text{ pc}^{-2} \text{ Gyr}^{-1},$$
also in this case, in agreement with observations. On the other hand, for the thick disc there is no constraint available, since there is no active star formation at the present time.

\subsection{Supernova rates}

\begin{figure}
\includegraphics[scale=0.4]{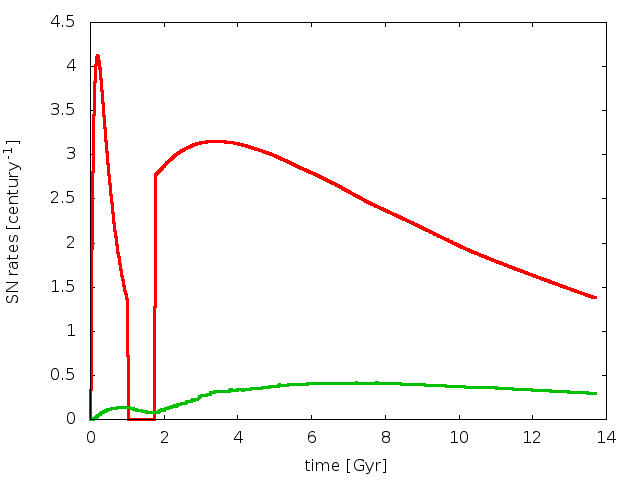}
 \caption{Temporal evolution of the SN rates, as predicted by model 2IM. \textit{Red line:} SNII rates predicted by model 2IM. \textit{Green line:} SNIa rates predicted by model 2IM.}
 \label{fig_09}

\includegraphics[scale=0.4]{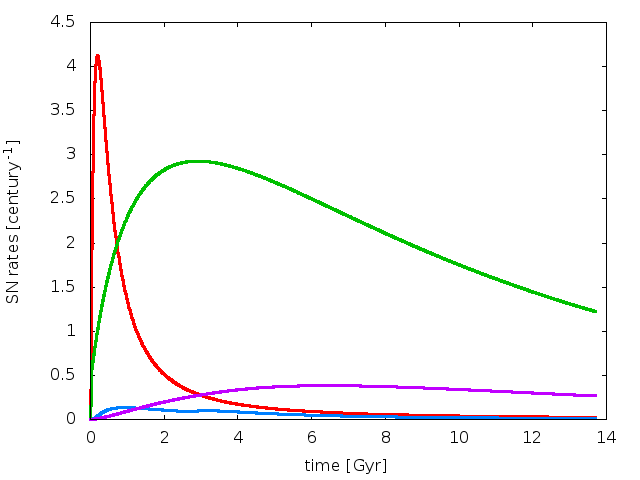}
 \caption{Temporal evolution of the SN rates. \textit{Red line:} SNII rates predicted by model 1IMT. \textit{Green line:} SNII rates predicted by model 1IMt. \textit{Blue line:} SNIa rates predicted by model 1IMT. \textit{Magenta line:} SNIa rates predicted by model 1IMt.}
 \label{fig_10}
\end{figure}

In Fig. \ref{fig_09}, we show the predicted behavior of the SN rates as a function of time as predicted by the two-infall model. As we can see, the gap between the end of the thick disc phase and the beginning of the thin disc phase, due to the adopted threshold, is also responsible for the trend of Type II SN rate. In fact, the trend shows a peak around $\sim$ 0.1 Gyr, which roughly corresponds to the timescale of formation of the thick disc phase, and then goes to zero at a time of about 1 Gyr, which corresponds to the end of the thick disc phase. The explanation of this feature is that the SNe Type II originate from stars with high mass and short lifetime, thus closely track the SFR and, hence, the number of this type of SNe per century is higher in the first gigayears of the formation of the Milky Way. Once the thick disc formation ends, star formation starts again and the number of supernovae per century increases until 3 Gyr, and then decreases until the achievement of the present rate. On the other hand, the SNe Type Ia are produced by progenitors with long lifetimes, thus they are very little influenced by the existence of a threshold in the star formation and the SNe Type Ia rate increases with time and remains almost constant until the achievement of the present value. An important constraint is represented by the present time SN rates in the solar vicinity, and according to Cappellaro and Turatto (1996) we have that:
$$ \text{SNII}=1.2\pm0.8 \text{ century}^{-1},$$
$$ \text{SNIa}=0.3\pm0.2 \text{ century}^{-1},$$
or more recently according to Li et al. (2011):
$$ \text{SNCC}=2.30\pm0.48 \text{ century}^{-1},$$
$$ \text{SNII}=1.54\pm0.32 \text{ century}^{-1},$$
$$ \text{SNIa}=0.54\pm0.12 \text{ century}^{-1}.$$
The values predicted by the model 2IM are respectively:
$$ \text{SNII}=1.4 \text{ century}^{-1},$$
$$ \text{SNIa}=0.3 \text{ century}^{-1},$$
in good agreement with observations.
\\On the other hand, in Fig. \ref{fig_10}, we show the predicted behavior of the SN rates as a function of time as predicted by the parallel model. 
The values of the present time SN rates in the solar vicinity predicted by the 1IMt are:
$$ \text{SNII}=1.2 \text{ century}^{-1},$$
$$ \text{SNIa}=0.3 \text{ century}^{-1},$$
also in this case, in agreement with observations.

\section{Conclusions}

In this work, we have studied the formation and evolution of the Milky Way thick and thin discs on the basis of detailed chemical evolution models to compare with the recent AMBRE [Mg/Fe] vs. [Fe/H] (Mikolaitis et al. 2017).
\\In particular, we have explored two different approaches for modelling Galactic chemical evolution: the two-infall and the parallel approach. In the two-infall scenario, the Galaxy formed by means of two infall episodes: during the first one, the thick disc formed whereas the second one gave rise to the thin disc. On the other hand, the parallel scenario assumes that the various Galactic components started forming at the same time but at different rates.
\\Our best models have been selected after performing several numerical simulations by varying one at the time the most important input parameters. The input parameters of the best models are: $\tau_1$=0.1 Gyr for the timescale of formation of the thick disc and $\tau_2$=7 Gyr for the timescale of formation of the thin disc, $\nu_1$=2 Gyr$^{-1}$ for the star formation efficiency of the thick disc and $\nu_2$=1 Gyr$^{-1}$ for the star formation efficiency of the thin disc, $\sigma_{th}$=7 M$_{\odot}$pc$^{-2}$ for the assumed threshold in the star formation rate, and a Kroupa et al. (1993) IMF.\\
\\Our conclusions are as follows.
\begin{itemize}
\item As regard to the abundance patterns, we have focused on the $\alpha$-element for which there is a clear distinction between thick and thin disc stars. The two-infall model can reproduce the thick and thin disc stars, but not the MRHA stars unless we assume that these stars have migrated from the inner thin disc. On the other hand, the parallel model treats the thick and thin discs as two truly distinct and parallel evolutionary phases and so we have two distinct tracks in the abundance pattern. With the parallel model, we are able to reproduce the MRHA stars as the metal rich thick disc stars, since in the parallel approach the thick disc can extend up to high [Fe/H], at variance with the two-infall sequential model.
\item For the metallicities distribution functions, the two-infall model can reproduce the MDF of the thick and thin disc stars, whereas it cannot reproduce the MRHA stars. On the other hand, with the parallel model the MDF of the thick disc is very broad and includes also the MRHA stars. We underline that the MDF represents a fundamental constraint for chemical evolution models because it is strongly dependent on the mechanism of disc formation. In particular, for our best models in the parallel scenario the timescale for the formation of the thick disc is equal to 0.1 Gyr, whereas the timescale for the formation of the thin disc at solar position is much longer and it is equal to 7 Gyr. Both these timescales are dictated by reproducing the MDF of each Galactic component.
\item Concerning the solar abundances, the predictions of all models are in reasonable agreement with the observations of Grevesse et al. (2007), but for Mg we had to increase the canonical yields from massive stars by a factor 1.2.
\item The predicted present-time SFR is $\psi_0=2.7 \text{ M}_{\bigodot} \text{ pc}^{-2} \text{ Gyr}^{-1}$ (for the two-infall model) and $\psi_0=2.4 \text{ M}_{\bigodot} \text{ pc}^{-2} \text{ Gyr}^{-1}$ (for the one-infall model of the thin disc), both in good agreement with observations. The predicted present-time SNII rate is 1.4 century$^{-1}$ (for the two-infall model) and 1.2 century$^{-1}$ (for the one-infall model of the thin disc), whereas the predicted present-time SNIa rate is 0.3 century$^{-1}$ (for the two-infall model) and 0.3 century$^{-1}$ (for the one-infall model of the thin disc), in good agreement with the observations.
\item In the two-infall approach, there is a gap in star formation between the thick and thin disc formation of several hundreds of Myr ($\sim$ 700 Myr), at variance with the parallel approach where no gap is present.
\end{itemize}
To summarize, a sequential approach like the one of the two-infall model can reproduce the chemical properties of thick and thin disc stars, but not those of the MRHA stars. In this case, these stars can be explained only by stellar migration from the inner disc. On the other hand, in order to reproduce the chemical properties of the MRHA stars without invoking stellar migration, it is better to consider a parallel scenario where the evolution of the thick and thin discs are separated. In this way, the MRHA stars can be interpreted as metal rich thick disc stars. However, the nature of these MRHA stars is still uncertain and forthcoming data concerning the ages of these stars will be fundamental to further constraint the disc formation and evolution.

\section*{Acknowledgments}

This work has made use of data from the European Space Agency (ESA) mission {\it Gaia} (\url{https://www.cosmos.esa.int/gaia}), processed by the {\it Gaia} Data Processing and Analysis Consortium (DPAC, \url{https://www.cosmos.esa.int/web/gaia/dpac/consortium}). Funding for the DPAC has been provided by national institutions, in particular the institutions participating in the {\it Gaia} Multilateral Agreement.
\\F.M., V.G. and E.S. acknowledge financial support from the University of Trieste (FRA2016). We also acknowledge useful discussions with A. Rojas-Arriagada.
\\Finally, we thank an anonymous referee for useful suggestions which improved the paper.

\end{document}